\documentclass[prb,showpacs,preprintnumbers,amsfonts,amssymb,amsmath,floats,twocolumn,aps,superscriptaddress]{revtex4}

\usepackage{graphicx}
\usepackage{dcolumn}
\usepackage{bm}
\usepackage[final,dvips]{epsfig}
\usepackage{color}

\newcommand{\tg}[0]{$t_{2g}$~}
\newcommand{\eq}[1]{Eq.~(\ref{eq:#1})}
\newcommand{\sct}[1]{Sect.~\ref{sec:#1}}
\newcommand{\fig}[1]{Fig.~\ref{fig:#1}}
\definecolor{darkgreen}{rgb}{0.0,0.5,0.0}  

\definecolor{funny}{rgb}{1.0,0.0,0.5}  

\newcommand{\ff}[1]{{\mathbf #1}}
\newcommand{\Tr}{\mbox{Tr}}

\begin{document} 
  
\title{
Momentum-resolved single-particle spectral function for TiOCl from
a combination of density functional and variational cluster calculations}

\author{M. Aichhorn}

\affiliation{
Insitut f\"ur Theoretische Physik und Astrophysik, Universit\"at
W\"urzburg, Am Hubland, 97074~W\"urzburg, Germany
}
\affiliation{
Centre de Physique Th\'{e}orique, \'{E}cole Polytechnique, CNRS, 91128
Palaiseau Cedex, France 
}

\author{T. Saha-Dasgupta}
\affiliation{
S.N. Bose National Centre for Basic Sciences, JD Block, Sector III,
Salt Lake City, Kolkata 700098, India
}

\author{R. Valent\'{\i}}
\affiliation{Institut f\"ur Theoretische Physik,
 Goethe-Universit\"at Frankfurt, 60438~Frankfurt/Main, Germany
}

\author{S. Glawion}
\author{M. Sing}
\author{R. Claessen}
\affiliation{Experimentelle Physik 4, Universit\"at W\"urzburg, Am
  Hubland, 97074~W\"urzburg, Germany} 

\begin{abstract}

We present results for the momentum-resolved single-particle
  spectral function of the 
low-dimensional system TiOCl in the insulating state, obtained by a
combination of {\it
  ab initio} Density Functional Theory (DFT) and Variational Cluster
(VCA) calculations. This approach allows to combine a realistic band
structure and a thorough treatment of the strong correlations. 
We show that it is important to include a realistic
{\em two-dimensional} band structure of TiOCl into the effective
strongly-correlated models in order to explain the spectral weight
behavior  seen in angle-resolved photoemission (ARPES)
experiments. In particular, we observe that the effect of the
interchain couplings is a 
considerable redistribution of the spectral weight around the $\Gamma$
point from higher to lower binding energies as compared to a purely
one-dimensional model treatment. Hence, our results support a description
of TiOCl as a two-dimensional compound with strong anisotropy and
also set a benchmark on the spectral features of correlated coupled-chain systems.

\end{abstract}


\maketitle

\section{Introduction}

In recent years a significant amount of research has been dedicated to
strongly-correlated materials with reduced dimensionality since
they exhibit 
a large variety of fascinating dimension-related properties.
 An example is the layered quantum
spin system TiOCl, where bilayers of Ti-O are separated by Cl$^-$
ions.  This system was originally  thought to be a possible candidate for a RVB
superconductor upon 
doping\cite{be.wi.93} because of its  frustrated triangular lattice
geometry. Later on, various experimental
 measurements\cite{se.ma.03,ru.ba.05,sh.sm.05,he.ho.05,sc.sm.06,ab.ma.07}
revealed that TiOCl  shows in fact an anomalous
spin-Peierls behavior with two consecutive 
phase transitions. Magnetic susceptibility
was initially described in terms of a one-dimensional spin-1/2 Heisenberg
model with a large intra-chain
coupling constant $J\approx 700$\,K.\cite{se.ma.03,ka.ba.03}.
It is though well known that susceptibility is not very sensitive
to different models and
recent {\it ab initio} DFT studies\cite{zh.je.08} showed that the underlying
interactions for this system can be understood in terms of  a spin-1/2 Heisenberg
model with strong intrachain antiferromagnetic interactions $J_1=660K$
and weaker interchain ferromagnetic interactions $J_2=-16K$, $J_3=-10K$. This
model reproduces the magnetic susceptibility measurements and sets a
framework for understanding the puzzling spin-Peierls phase
transitions in this compound. Only recently, research has also focused
on high-pressure studies\cite{ku.fr.06,ku.pa.08,zh.je.08.2,bl.ri.08u}
as a possible way to drive the system metallic.

At room temperature and ambient pressure, the system is a Mott
insulator with a charge gap 
of about 2\,eV.\cite{ru.ba.05,ku.fr.06} The electronic structure in
this high-temperature phase has been examined by angle-resolved
photoemission spectroscopy (ARPES).\cite{ho.si.05,ho.si.07} In agreement
with previous experimental evidence, the results show a strong
anisotropy of the correlated band structure, with significant
dispersion of the Ti $3d$ bands along the chains (crystallographic
$b$-direction), and  almost flat bands perpendicular to the chains.

On the theoretical side, the electronic properties of TiOCl have been
studied by means of {\it ab initio} DFT calculations within the local density
approximation (LDA), the 
LDA+U \cite{se.ma.03,da.va.04}, B3LYP \cite{pi.va.07}
and also in combination with
the dynamical mean-field theory (DMFT),\cite{da.li.05,cr.la.06,da.li.07} 
which is a modern method for dealing with strong correlations. 
It was shown that a proper treatment of non-local
correlations is crucial for a reasonable description of the
single-particle gap.\cite{da.li.07} 

However, the momentum dependence of the spectral function $A(\mathbf
k,\omega)$ seen in
ARPES is still puzzling. It has been shown that an {\em ab initio}
calculation without proper treatment of correlations is
insufficient.\cite{se.ma.03,da.va.04,pi.va.07,ho.si.07} On the other
hand, describing the compound by a simplified one-dimensional strongly 
correlated model was not successful either.\cite{ho.si.05}
Furthermore, LDA+DMFT could so far only
produce the momentum integrated local density of states
(DOS) without any information on the momentum dependence of the
spectra.\cite{da.li.05,cr.la.06,da.li.07} 
This situation, having no calculation for the momentum-resolved
spectral function $A(\mathbf k,\omega)$ at
hand, is partly due to the fact that there are only few methods that
can deal with all the requirements of such calculations. 
This work is intended to fill this gap and investigates the
influence of the true two-dimensional band-structure on the
momentum-resolved $A(\mathbf k,\omega)$ in the presence of strong
correlations. A successful technique for this purpose is the Variational
Cluster Approach (VCA)\cite{po.ai.03,pott.03.epjb2}.

In what follows we apply a two-step procedure to study the spectral
function, as has been proposed by Chioncel {\em et
  al.}.\cite{ch.al.07} First, DFT calculations within the LDA are
carried out, and localized Wannier functions are constructed by the
N-th order muffin-tin-orbital (NMTO)\cite{an.da.00} downfolding technique.
Using the LDA Hamiltonian expressed in these Wannier functions as the
non-interacting part, and adding Coulomb and Hund interaction terms,
we arrive at the correlated low-energy model.
By applying VCA to this model Hamiltonian,
we show that the inclusion of the
inter-chain processes leads to a significant redistribution of spectral
weight from higher to lower binding energies.
Since these processes enhance the asymmetry of the strongly-correlated
band structure, they
are crucial for the reproduction of the asymmetric
bands seen in ARPES measurements. Our calculations show that the
Hubbard-model description is appropriate for TiOCl if effects
beyond the one-dimensional description are included.  Moreover
these results should be valid for a large variety of correlated low-dimensional
coupled-chain systems.

The paper is organized as follows: In \sct{theory} we discuss the
construction of the low-energy Hamiltonian, as well as the VCA, which
is subsequently used
for the calculation of the correlated spectral function. \sct{results}
contains our results of the multi-band as well as of the single-band Hubbard
model and in \sct{discuss} we present our discussions
and conclusions.

\section{Theory}\label{sec:theory}

In many transition metal oxides electronic correlation effects are
very important for a proper description of the physical
properties. However, it is a known fact that first-principle
calculations suffer from an insufficient treatment of these
effects. In order to take the strong correlations into account in our
calculation for TiOCl, we
apply a two-step procedure ({\em LDA+VCA}) that has first been introduced by
L.~Chioncel {\em et al.}.\cite{ch.al.07} It consists of the
construction of the correlated low-energy Hamiltonian based on
density-functional 
theory on the one hand, and the solution of the resulting low-energy
Hamiltonian 
using the VCA on the other hand. In Ref.~\onlinecite{ch.al.07}, the
authors study the non-quasiparticle states in the half-metallic
compound CrO$_2$ and find good agreement with experiments. Moreover, a
comparison with LDA+DMFT calculations showed the applicability of the
LDA+VCA approach. Recently, it has also been used to explain the
pseudogap in TiN, where also the momentum-resolved spectral function
has been calculated.\cite{al.ch.09}

\subsection{Low-energy Hamiltonian}

For a complete description of the electronic structure of a given
material it is in principle necessary to consider all electronic
degrees of freedom of the underlying constituents. Calculations within
DFT can to some extent fulfill this
requirement. However, it is clear that only certain
states and orbitals contribute to the low-energy physics. For
this reason one is interested in finding an effective model that
describes the low-energy physics on the one hand sufficiently accurate
and has, on the other hand, not too many degrees of freedom. 

In the present case of TiOCl, DFT calculations within the LDA
approximation have shown that the relevant orbitals at low energies
are the Ti 3d orbitals, which are split into \tg and $e_g$
manifolds due to the octahedral crystal field provided by the
ligands. Since the Ti$^{3+}$ ion is in a 3d$^1$ configuration, 
the relevant states closest to the Fermi energy are of
  predominantly \tg character.

For the construction of the low-energy Hamiltonian, we performed DFT
calculations within the LDA using the linearized muffin-tin-orbitals
(LMTO) basis set. The localized orbitals, which are the basis of the
interacting model, are constructed using the NMTO method. By using the
downfolding technique\cite{an.da.00}, the hybridization of the Ti-\tg
orbitals with the ligand orbitals (O-$p$ and Cl-$p$) are taken into
account, yielding an effective set of \tg orbitals. These orbitals
represent the LDA band structure with great accuracy, and are used as
the non-interacting part of the many-body Hamiltonian. The matrix elements
of the NMTO Hamiltonian $H^{\rm LDA}(\mathbf k)$ in the basis set of
localized NMTO Wannier functions give
the transfer integrals $t_{ij}^{\alpha\beta}$, and the non-interacting
Hamiltonian can be written as
\begin{equation}
  H_0^{\rm LDA} =
  \sum_{ij,\sigma}\sum_{\alpha\beta}t_{ij}^{\alpha\beta}c^\dagger_{i\alpha,\sigma}c^{\phantom{\dagger}}_{j\beta,\sigma}.
\end{equation}
The indices label the lattice sites by $i$,~$j$, as well as the \tg
orbitals by $\alpha$,~$\beta$, and $\sigma$ denotes the spin.

To include correlation effects into the low-energy
description, we add 
interaction terms to the Hamiltonian,
\begin{equation}\label{eq:ham}
  \begin{split}
    H=&H_0^{\rm LDA}-\mu\sum_{i\alpha}n_{i\alpha}+\frac{U}{2}\sum_{i\alpha\sigma}n_{i\alpha\sigma}n_{i\alpha\bar\sigma}\\
    &+\frac{U^\prime}{2}\sum_{i,\alpha\neq\beta}n_{i\alpha}n_{i\beta}-J^z\sum_{i,\alpha\neq\beta}S_{i\alpha}^z S_{i\beta}^z\\
    &-\frac{J}{2}\sum_{i,\alpha\neq\beta}\left(S_{i\alpha}^+ S_{i\beta}^- + S_{i\alpha}^- S_{i\beta}^+\right).
  \end{split}
\end{equation}
For convenience, we introduced the chemical potential $\mu$ in the Hamiltonian.
We will refer to this Hamiltonian as  \tg model, and give all energies
throughout the paper in units of electron volt (eV). 
The full low-energy model
\eq{ham} consists of the single-particle terms ($H_0$ and $\mu$), 
the diagonal (density-density) interactions ($U$, $U^\prime$, and $J^z$), 
and the non-diagonal (spin-flip) term ($J$, third line). In this study
we consider only the case $J^z = J$. $n_{i\alpha}= n_{i\alpha\uparrow}+n_{i\alpha\downarrow}$ 
is the orbital occupation operator, and $S_{i\alpha}^z$,
$S_{i\alpha}^+$, $S_{i\alpha}^-$ are the components of the
spin-$\frac{1}{2}$ operator on site $i$ in orbital $\alpha$. 
The interaction parameters $U$, $U^\prime$, and $J$ are not independent, but
fulfill the relation $U^\prime=U-2J$. At this point it is important to
note that we can include a full $SU(2)$ symmetric exchange term. Since
the method we consider is not affected by any sign problem, it has no
restriction
 on the type of couplings that can be
included.\cite{qmc} For the interaction parameters $U$ and $J$ one can
find several values in the literature, ranging from $U=3.0$\,eV to
$U=4.0$\,eV and $J=0.5$\,eV to
$J=1.0$\,eV.\cite{se.ma.03,da.va.04,da.li.05,ho.si.05,da.li.07} Since
we 
want to study also the influence of these parameters on the
single-particle properties, we have performed calculations
with  different values, and
indicate the actual value at the corresponding location in the
paper.

In this work, we also address the question whether the orbital degrees of
freedom are important for the low-energy physics or not. Since TiOCl
does not crystallize in a perfect cubic symmetry, the threefold
degeneracy of the \tg manifold is
lifted. LDA+U\cite{se.ma.03,da.va.04} and also LDA+DMFT\cite{da.li.05}
calculations have shown that the ground state shows predominantly
d$_{xy}$ character (the local reference frame is $\hat z=a$, and $\hat
 x$ and $\hat y$ axes rotated by 45$^\circ$ with respect to $b$ and $c$),
 with only very small admixture of the other 
orbital degrees of freedom, a picture that we  will also find in our following
 calculations. 
This is in contrast to IPT-DMFT
calculations,\cite{cr.la.06} where a sizable admixture of the other
orbital degrees of freedom is found.

In order to investigate the effective one-band model that consists of
the d$_{xy}$ orbital only, we performed a NMTO downfolding procedure
integrating out all other degrees of freedom, and keeping only the
d$_{xy}$ channel. In this one-band model, the only interaction terms
are the ones proportional to the Hubbard onsite $U$, and the
low-energy one-band Hamiltonian finally reads
\begin{equation}\label{eq:ham1b}
  H=\sum_{ij,\sigma}t_{ij}c^\dagger_{i\sigma}c^{\phantom{\dagger}}_{j\sigma}
  +U\sum_in_{i\uparrow}n_{i\downarrow} - \mu\sum_{i}n_{i}\; ,
\end{equation}
where the Hubbard interaction $U$ is the same as for the \tg-model.
The effective hopping parameters $t_{ij}$ are again the matrix
elements of $H^{\rm LDA}(\mathbf k)$ in the Wannier basis set.

\subsection{Variational Cluster Approach}

\begin{figure}[t]
  \centering
  \includegraphics[width=0.8\columnwidth]{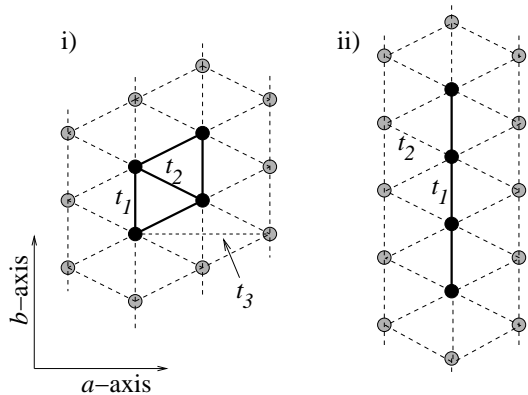}
  \caption{\label{fig:cluster}%
    Triangular lattice structure and two possible clusters tiling the
    lattice. i) 4-site 
    cluster including inter-chain self-energies. ii) 1D clusters
    neglecting inter-chain self-energies. Full circles and solid lines
    mark sites and bonds inside a clusters. Next-next-nearest neighbor
    hopping $t_3$ is only drawn once for clarity. In the case of the
    \tg manifold each lattice site consists of three orbitals.
  }
\end{figure}

After having constructed the low-energy Hamiltonian using {\em ab
  initio} techniques, we use the VCA\cite{po.ai.03,pott.03.epjb2}
in order to calculate the spectral function of this model. Since we
deal with an effective low-energy Hamiltonian that involves no other
uncorrelated ligand states (O-$p$, Cl-$p$), but only the correlated
Ti-\tg orbitals, the application of VCA is
straightforward, and is from a technical point of view exactly
equivalent to standard multi-orbital calculations for Hubbard-model
Hamiltonians. The only difference is that the non-interacting part is
determined by the procedure discussed in the previous subsection.
Furthermore, since there are no explicit ligand states in the
Hamiltonian, there is no need for a double-counting 
correction. It gives just a constant shift in energy which can be
absorbed in the chemical potential.

As mentioned in the introduction, the VCA is a quantum cluster method
capable of treating strong short-ranged correlations.
The main idea is to approximate the self-energy of the
original model, which is defined on an infinite lattice, by the
self-energy of a finite cluster, the reference system. The 
variational principle states that the optimal solution is given by the
stationary points of the grand potential
$\Omega[\Sigma]$ as a function of the self-energy $\Sigma$. Parametrizing the self-energy by the single-particle
parameters $\ff t^\prime$ of the reference system, we can write the
grand potential as 
\begin{equation}\label{eq:ocalc} 
        \Omega (\ff t^\prime) 
        =\Omega^\prime + \Tr \ln (\ff G_{0,\ff t}^{-1}-\ff \Sigma(\ff t^\prime))^{-1}
        - \Tr \ln \ff G_{\ff t^\prime} \: ,
\end{equation}
where $\Omega^\prime$  and $\ff G_{\ff
  t^\prime}$  are respectively the grand canonical potential and
the Green's function  of the reference system and $\ff
G_{0,\ff t}$ is the non-interacting Green's function of the physical
(lattice) system. The stationary condition reads 
\begin{equation}
  \frac{\partial \Omega}{\partial \ff t^\prime}\Big|_{\ff t^\prime=\ff t^\prime_{\rm opt}} = 0 \; .
\end{equation}

It is important to note that the interaction parameters are not
variational parameters, since, by construction of the VCA, the
interaction terms of the reference system and the original lattice
model must not differ. In this study, we restrict ourselves to local
interactions only since the VCA in its strict sense cannot be used
for models with non-local interactions without further
approximations. 

In its spirit, the VCA is closely related to the dynamical
  mean-field theory (DMFT), where in the latter case the self-energy
  is obtained from an impurity problem. 

The actual VCA calculation is done in the following steps. First,
  we determine the ground state of the reference system,
  i.e. a cluster of finite size as depicted in \fig{cluster}.
  The interacting
  Green's function is calculated, and since the non-interacting
  Green's function of the 
  reference system is known, the self-energy can readily be obtained
  using Dyson's equation. Using the grand potential $\Omega^\prime$ of
  the reference system, the Green's function $\ff G_{\ff
    t^\prime}$, and the self-energy $\ff \Sigma(\ff t^\prime))$,
  Eq.~\ref{eq:ocalc} is evaluated using the technique of
  $Q$-matrices.\cite{ai.ar.06.2} Note that the Green's functions $\ff
  G_{0,\ff t}$, $\ff G_{\ff t^\prime}$ and the self-energy
  $\ff\Sigma(\ff t^\prime)$ in \eq{ocalc} are matrices not only in
  site and spin indices, but also carry an orbital index. In fact,
  this is the only difference of the application of VCA in the present
  case compared to the numerous previous applications to the
  single-band Hubbard model. 

As a reference system solver, we
use the Band-Lanczos exact diagonalization technique at zero
temperature, which means that for the full \tg manifold, we can 
easily consider clusters with at most 4 sites, yielding a 12-orbital 
Green's function $\ff G_{\ff t^\prime}$.
For the single-band model, we consider clusters up to 12
sites. 
We exploit particle number and spin conservation, therefore the sizes
of the largest Hilbert spaces that we have to  
consider are $N=14520$ states in the 4-site multi-orbital case, and
$N=853776$ states in the 12-site single-orbital case, respectively.
Since we are considering an exact diagonalization method for solving the
cluster problem, all interactions in the Hamiltonians \eq{ham} and
\eq{ham1b} are treated exactly and on the same footing. This is a
clear advantage compared to, e.g., using the Hirsch-Fye quantum
Monte-Carlo method as impurity solver, since in the latter case
approximations to the interaction terms of the Hamiltonian have to be
done.\cite{qmc}

The VCA approach has been
tested thoroughly and used successfully for many investigations in
recent years. 
Several studies on the cuprate-based high-temperature
  superconductors have shown that this approach can reproduce salient
  features of these materials, such as the ground-state phase
  diagram,\cite{se.la.05,ai.ar.05,ai.ar.06,ai.ar.06.2}
  or the opening of the pseudogap at
  low hole doping, accompanied with the occurrence of Fermi
  arcs,\cite{se.tr.04,ai.ar.06} in very good agreement with
  experiments and results
  obtained by the cellular dynamical mean-field theory (CDMFT) (see, e.g,
  Refs.~\onlinecite{tr.ky.06}). Recently, the VCA could also reproduce
  the pairing symmetry of the iron-based
  superconductors.\cite{da.mo.08}  

The VCA has also been used for multi-orbital systems, which is
  relevant for the combination with {\em ab-initio} methods. On the
  pure methodological level, the metal-insulator transition in infinite
  dimensions was studied\cite{in.ko.05,in.ko.07}, and very good
  agreement with dynamical mean-field calculations was found. An
  application to real materials was done in
  Refs.~\onlinecite{ed.07,ed.08}, where the compounds NiO, CoO, and
  MnO have been studied and very good agreement with experimental
  photo-emission data has been found.

Details on the practical implementation of the VCA, including tests
and benchmarking, can be found in
Refs.~\onlinecite{po.ai.03,pott.03.epjb1,po.05.ass,ai.ar.06.2,ba.ha.08,se.08u}.

In general, all the single-particle parameters $\ff t^\prime$ are
variational parameters of the VCA. In practice, one chooses a
physically motivated subset in order to keep the numerical
calculations feasible. Here we make the following choice.
For a thermodynamically consistent description of the densities,  it is
crucial to consider the onsite energies, i.e. the local terms of the
single-particle Hamiltonian 
$\varepsilon_\alpha^\prime \equiv
(t_{ii}^{\alpha\alpha})^\prime$, as variational 
parameters.\cite{ai.ar.06}
We define  
the average $\varepsilon^\prime =
\frac{1}{2}(\varepsilon_{xy}^\prime+\varepsilon_{yz}^\prime)$ and the
crystal-field splitting $\Delta_{\rm cf}^\prime =
\varepsilon_{yz}^\prime - \varepsilon_{xy}^\prime$, which are then
used as the 
variational parameters of the VCA. Note that in the single-band case,
\eq{ham1b}, one has to deal with $\varepsilon^\prime$ only. One has to be
aware that the variational parameter $\Delta_{\rm cf}^\prime$ {\em 
    does not} impose an artificial orbital polarization of the
  system, since it is a parameter in the variational procedure and no
  physical external field. Hence, using $\varepsilon^\prime$ and
  $\Delta_{\rm cf}^\prime$, the orbital occupancies are determined
  in a fully self-consistent way.

The main property investigated in this work is the single-particle
spectral function which we define as 
\begin{equation}\label{eq:akw}
  A(\ff k,\omega) = -\frac{1}{\pi}{\rm \:tr\:Im\:} {\ff G}(\ff k,\omega) \;,
\end{equation}
Since we broke the translational invariance of the system by
  introducing the cluster tiling, a proper periodization of the
  lattice quantities is needed in order to restore translational
  symmetry, an issue also important in cluster DMFT
  calculations.\cite{bi.pa.04} Here, we choose to use the
  periodization of the Green's function since the periodization of the
  self-energy gives unphysical results in the insulating phase.\cite{se.08u}
  In other words, starting from the Green's function that depends on
  two momenta, ${\ff G}(\ff k,\ff k^\prime,\omega)$, one restores the fully 
  translationally invariant Green's function ${\ff G}(\ff k,\omega)$,
  by neglecting the off-diagonal elements, and taking $\ff k=\ff k^\prime$
  only. It has been shown that this is a well justified approximation
  to calculating the momentum-dependent spectral function.\cite{se.pe.00}
The Green's function $\ff G$ is in general a matrix in orbital
indices and $A(\ff k,\omega)$ is given by the trace over the orbital
degrees of freedom.

\section{Results}\label{sec:results}

\subsection{Full \tg model vs. effective one-band model}\label{sec:t2g}

\begin{figure}[t]
  \centering
  \includegraphics[width=0.8\columnwidth]{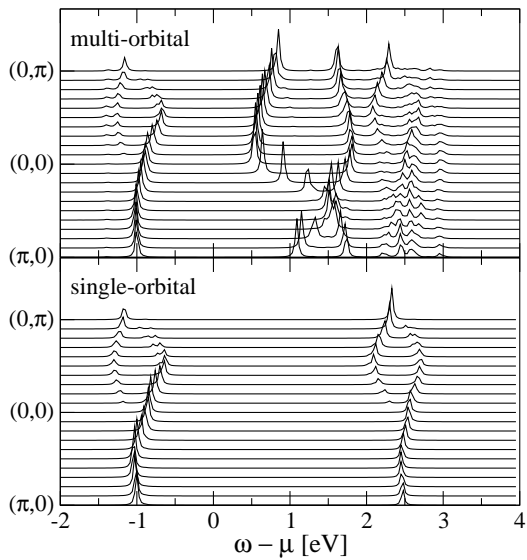}
  \caption{\label{fig:comp_mo_so}%
    Comparison of the single-particle spectral function of the \tg
    model (top) and the single-band model (bottom), both calculated
    with a 
    $2\times 2$ reference system, see \fig{cluster} i).
    Parameters
    are $U=3.3$\,eV, $J=0.5$\,eV. For the hopping parameters see
    text. The chemical potential has been chosen such that i) the
    system is insulating with $n=1.0$, and ii) the position of the
    occupied states coincide in both calculations. Lorentzian
    broadening of $\eta=0.02$\,eV was used.
  }
\end{figure}
\begin{figure}[t]
  \centering
  \includegraphics[width=0.8\columnwidth]{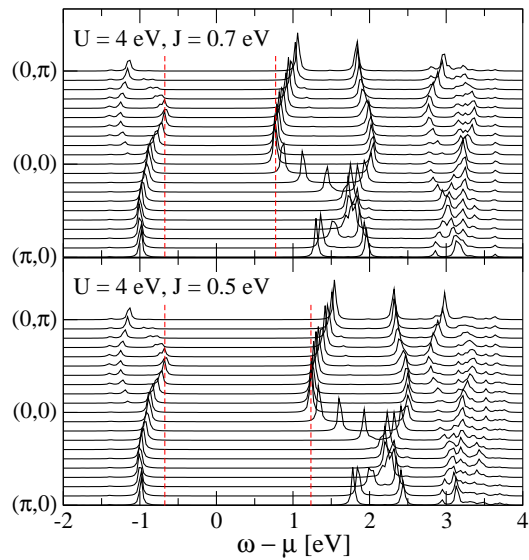}
  \caption{\label{fig:comp_U_J}%
    (Color online) Comparison of the spectral function of the \tg
    model for two different sets of interaction parameters. Top:
    $U=4.0$\,eV, $J=0.7$\,eV. Bottom: $U=4.0$\,eV, $J=0.5$\,eV. The
    vertical dashed lines mark the edges of the single-particle gap. Lorentzian
    broadening of $\eta=0.02$\,eV was used.  
  }
\end{figure}

Before we come to a detailed analysis of the single-band model, we
first want to check the validity of the restriction to the lowest
$d$-orbital.

For this reason, we
performed {\it ab initio} calculations to determine the full single-particle
Hamiltonian of the system. Since Ti$^{3+}$ is in a $3d^1$
configuration, the $e_g$ orbitals are unoccupied and can be projected
out, and the full kinetic Hamiltonian is downfolded to the threefold
degenerate \tg manifold. Results show, that the system exhibits a
strong anisotropy, with the largest hopping integrals in the
crystallographic $b$-direction, see \fig{cluster}, almost one
order of magnitude larger than the other transfer integrals. Moreover,
the threefold degeneracy is lifted and the manifold split into the
lower $d_{xy}$ and the higher $d_{xz}$ and $d_{yz}$ orbitals.
 Note that for the orbital
designation we consider
the same local reference frame as in Ref.~\onlinecite{da.li.05,ho.si.05}.  The
crystal-field splitting between the ground state and the
first excited state obtained from LDA is about 0.42\,eV. This theoretical
value is in reasonable agreement with experimental
results.\cite{ru.ba.05,za.de.06}  
Despite this splitting, the orbital sector is not
fully polarized in the LDA calculations, and the occupation of
$d_{xy}$ is about 0.49, with 0.51 electrons in the other two
orbitals. 

In order to perform our LDA+VCA calculations, we take the downfolded
Hamiltonian of the {\it ab initio} calculations, and add the interaction and
exchange terms according to \eq{ham}. In the upper panel of
\fig{comp_mo_so} we show the results for the spectral function,
\eq{akw}, calculated for this three-band \tg model using typical
parameters $U=3.3$\,eV and $J=0.5$\,eV. The bands which
are located just above the Fermi level, between roughly 0.5\,eV and
2.0\,eV, have $d_{xz}$ and $d_{yz}$ character, and remain almost
unchanged by the strong interactions. 

The behavior of the $d_{xy}$
orbital is strikingly different. It splits into two bands that can be
identified with the lower and upper Hubbard band, located roughly around
$-1.0$\,eV and 2.5\,eV, respectively. 
By inspecting the terms of the Hamiltonian related to the crystal-field
splitting, i.e.
$H_{\rm cf} = \Delta_{\rm cf} \sum_i (n_{yz}+n_{zx}-n_{xy}) $, we can
calculate the orbital polarization $p=\partial \Omega / \partial
\Delta_{\rm cf}$. We find a value of $p=-0.99$, meaning that the
system is almost perfectly polarized into the $d_{xy}$ orbital, which
is in agreement with recent LDA+CDMFT calculations.\cite{da.li.07}
Interestingly, this polarization is found without any sizable increase
of the crystal-field splitting in the variational procedure,
i.e., $\Delta_{\rm cf}\approx \Delta_{\rm cf}^\prime$, but it is only due
to the inclusion of strong local interactions.

This result gives rise to the question, to which extent a single-band
Hubbard model can describe the occupied states relevant for
comparison with ARPES. We took parameters from a full downfolding to
the Ti $d_{xy}$ orbital only, cf. first column of Table~I in
Ref.~\onlinecite{da.va.04}. Using the same value of $U=3.3$\,eV we
calculate the spectral function for Hamiltonian \eq{ham1b}.  The results 
 are shown in the lower panel of \fig{comp_mo_so}. In order to
avoid effects coming from different cluster sizes, we
used also a $2\times 2$ cluster for this comparison. Note
that  below the Fermi level the agreement between
the single-band model and the $d_{xy}$ part of the \tg model is
excellent. For this reason we conclude that for a comparison of
spectra with experimental ARPES measurements the Hamiltonian \eq{ham1b}
is a reasonable starting point.

Before turning to a more detailed analysis of the spectra obtained from 
\eq{ham1b}, let us briefly discuss the single-particle gap
$\Delta$, defined as the energy difference between the lowest
  unoccupied and the highest occupied state.
For a comparison of this 
quantity with experiments, it is clear that the single-band model is
not sufficient, since it does not describe the excited states in the
\tg manifold. However, we extracted the gap $\Delta$ from the spectral
function of the \tg model, for different values of
$U$ and $J$. We find that the main quantity that determines the gap is
the inter-orbital Coulomb interaction $U^\prime=U-2J$, and 
we get $\Delta\approx
1.2$\,eV for $U=3.3$\,eV, $J=0.5$\,eV. In \fig{comp_U_J} we plot the spectral
function for different sets of interaction parameters, and find
$\Delta\approx 1.4$\,eV for $U=4$\,eV, $J=0.7$\,eV, and $\Delta\approx
1.9$\,eV for $U=4$\,eV, $J=0.5$\,eV. All these values for the gap are
a bit smaller than the experimental charge gap of about
2\,eV,\cite{ru.ba.05} but nevertheless in reasonable agreement.

\subsection{Spectral weights in the single-band Hubbard model}

\begin{figure}[t]
  \centering
  \includegraphics[width=0.95\columnwidth]{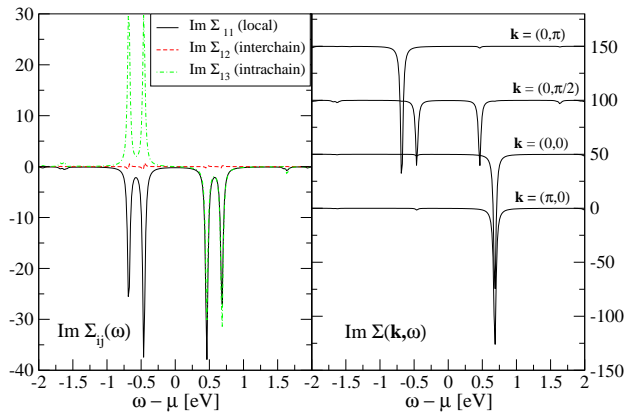}
  \caption{\label{fig:selfenergy}%
    (Color online) Imaginary part of the self-energies on a reference
    system consisting of two coupled 4-site chains. Left panel: Real
    space $\Sigma_{ij}(\omega)$. 
    Solid line: Local/onsite self-energy $\Sigma_{11}(\omega)$.
    Dashed line: Inter-chain self-energy $\Sigma_{12}(\omega)$. 
    Dash-dotted line: Intra-chain self-energy $\Sigma_{13}(\omega)$. 
    Right panel: Self-energy for selected momenta of the
    cluster. Momenta are indicated in the plot, and a
    vertical shift between momenta has been introduced for improved
    presentation. The self-energy
    shows causality (negative definite), almost no dependence in
    $a$-direction, and strong dependence in $b$-direction. Lorentzian
    broadening of $\eta=0.02$\,eV was used in both plots.
  }
\end{figure}

We have shown in Sec.~\ref{sec:t2g} that the occupied states of the
\tg manifold are well reproduced by a single-band Hubbard model. In
this section we want to investigate the spectral function of
\eq{ham1b} in more detail. As mentioned above, we focus 
on the effect of the additional two-dimensional hopping processes  
on the quasi-one-dimensional behavior of TiOCl. 

First, we want to determine the strength of the correlations along the
qualitatively different bonds of the lattice, \fig{cluster}. This
can  be done best by inspecting the self-energy $\Sigma_{ij}(\omega)$,
which is, in VCA, a quantity defined on the reference system and, thus,
can be readily obtained. The results for three selected matrix
elements are shown in \fig{selfenergy}, calculated on an $2\times 4$
cluster. 
This cluster consists of two coupled 4-site chains in
  $b$-direction. In other words, a cluster similar to the one depicted
  in \fig{cluster} i) but with doubled extension in $b$-direction.
It is obvious that local ($\Sigma_{11}(\omega)$)
and intra-chain correlations ($\Sigma_{13}(\omega)$) are strong, but
the correlations between adjacent chains ($\Sigma_{12}(\omega)$) are
orders of magnitude weaker. 
In addition, we show in the right panel of \fig{selfenergy} the
  Fourier transformation of the self-energy, for momenta accessible at
  this small cluster. Again, the influence of momenta perpendicular to
  the chains is hardly visible in the self-energy, whereas it shows
  significant momentum dependence in the chain direction.
This leads to the conclusion that the
spectra of TiOCl should be governed by 1D correlations, modified by
single-particle effects due to the coupling of the chains.

\begin{figure}[t]
  \centering
  \includegraphics[width=0.8\columnwidth]{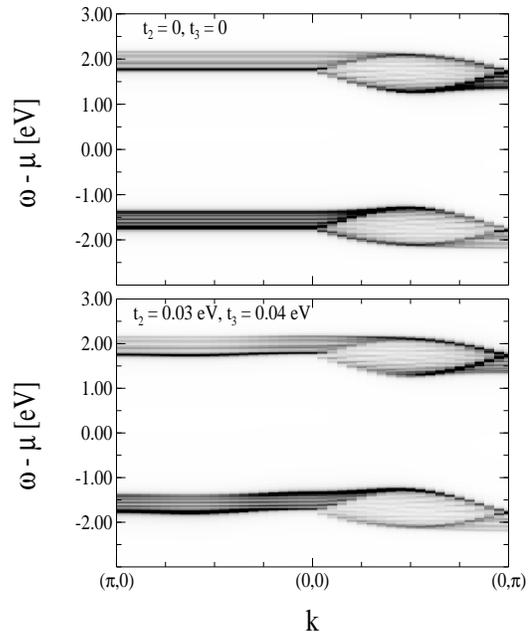}
  \caption{\label{fig:comp_singleband}%
    Spectral function of the single-band model with
    a 12-site chain as reference system. Top: No coupling between
    chains. Bottom: Inclusion of interchain coupling parameters obtained from LDA-downfolding. Dark areas
    mark large spectral weight. Lorentzian
    broadening of $\eta=0.02$\,eV was used. 
    The horizontal striped
    structures occur due to the breaking of translational invariance
    in the cluster approximation.
  }
\end{figure}

Motivated by this result, we use from now on a $1\times 12$ cluster as
reference system. Since the VCA approximates the interacting Green's
function as ${\mathbf G}(\omega)^{-1} = \ff G_{0,\ff t}^{-1}-\ff
\Sigma(\omega)$ with $\ff G_{0,\ff t}$ the non-interacting Green's
function of the model Hamiltonian and $\ff \Sigma(\omega)$ the cluster
self-energy, it is easy to see that the inter-cluster coupling is
treated in a single-particle (i.\,e., non-interacting) manner, since
it enters just via $\ff G_{0,\ff t}$. 
On the other hand, this procedure gives the best possible description
of the correlation effects along the chains in $b$-direction.

\begin{figure}[t]
  \centering
  \includegraphics[width=0.8\columnwidth]{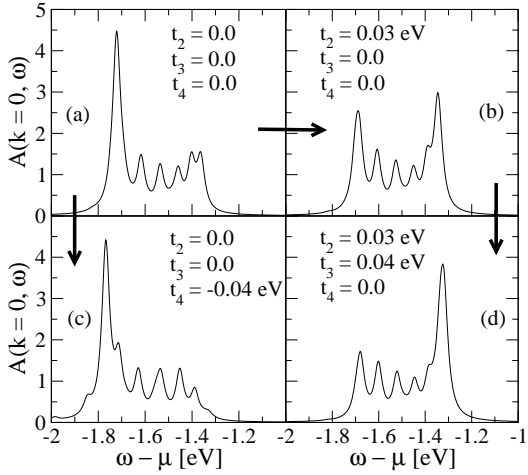}
  \caption{\label{fig:comp_Ak0}%
    Spectral function $A(k=0,\omega)$ at the $\Gamma$ point. Top left:
    without inter-chain coupling. Top right: Including nearest-neighbor
    inter-chain coupling $t_2$. Bottom right: Including
    next-nearest-neighbor inter-chain coupling $t_3$. For comparison,
    bottom left: next-nearest-neighbor hopping along the chain, but no 
    inter-chain hopping. Lorentzian
    broadening of $\eta=0.02$\,eV was used.
}
\end{figure}

In \fig{comp_singleband} we show a density plot of the spectral
function of the Hamiltonian \eq{ham1b} for $U=3.3$\,eV. In the upper
part we included only the intra-chain hopping $t_1=-0.21$\,eV in the
calculation, leading to flat bands in $a$-direction, i.\,e., from
$(\pi,0)$ to $(0,0)$, since in this case the chains are
  decoupled. By including additional inter-chain parameters 
$t_2=0.03$\,eV and $t_3=0.04$\,eV as given in Ref.~\onlinecite{da.va.04}, we
notice a slight dispersion in $a$-direction. In $b$-direction,
however, the band positions remain almost unchanged; we find only
changes in the spectral weights. Since this cannot be seen clearly in
the density plots, we show in \fig{comp_Ak0} the evolution of the
spectral function at the $\Gamma$ point $\mathbf k=(0,0)$ when
longer-ranged hopping processes are included. 

The upper left panel (a) is the spectral function for decoupled chains
with the spin-charge separation clearly visible. At the $\Gamma$ point
the holon band is
located around $-1.72$\,eV, and the spinon band around
$-1.39$\,eV. Including the nearest neighbor inter-chain hopping
$t_2$ leads to a significant
redistribution of spectral weight from the holon to the spinon
band, i.\,e., from higher to lower binding energies, see the upper right panel (b).
This effect is even enhanced when the next-nearest inter-chain
hopping $t_3$ is included, as shown in the lower right panel (d). Here
the spectral weight of the low binding energy ('spinon') excitation is
comparable to the weight of the high binding energy ('holon')
excitation for decoupled chains in panel (a), and vice verse.
At this point we would like to mention, that in a strict sense the
terminology 'spinon' and 'holon' is not applicable any more, since
these are properties of purely one-dimensional systems. Anyway, since
the spectra resemble to some extent 1D systems, we still use these
terms to distinguish the different excitations.

From \fig{comp_Ak0} it is clear that the inclusion of
inter-chain processes enhances the asymmetry of the $\mathbf
k$-resolved spectra.  The  low-lying excitation near the $\Gamma$
point is strongly enhanced, whereas there is no spectral weight
transfer to lower binding energies visible around $(0,\pi)$. Note that
we define the spectrum to be symmetric if the main excitations at
$\mathbf k$-vector $(0,0)$ and $(0,\pi)$ are located at the same
binding energy.

One may ask if it is possible to get
 a similar spectral weight distribution by using
only the purely one-dimensional Hubbard model, but including
longer-ranged intra-chain hopping processes as given by the {\it ab initio}
calculations. In fact, the next-nearest-neighbor hopping term along
the chain, $t_4$, is of similar size of the inter-chain
hoppings.\cite{da.va.04} The result for the spectral function at the
$\Gamma$ point in this pure 1D case is shown in the lower left panel (c) of
\fig{comp_Ak0}. 
From this result it is obvious that one can not get an
excitation at binding energies of roughly
$-1.4$\,eV, as seen in experiments. On the contrary, the spectral
weight of the excitation at the higher binding energy of about
$-1.8$\,eV is even enhanced in the one-dimensional treatment
when longer-ranged hopping processes are included.

\begin{figure}[t]
  \centering
  \includegraphics[width=0.9\columnwidth]{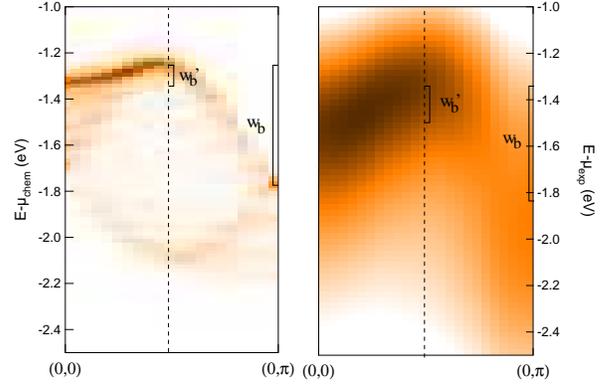}
  \caption{\label{fig:comp_exp}%
    (Color online) Comparison of the theoretical spectral function (left) with the
    experimental ARPES spectra (right). Only the
    $b$-direction is shown. Lorentzian broadening of $\eta=0.02$\,eV
    was used. 
    The vertical dashed line marks momentum $(0,\pi/2)$, and
    the band widths $w_b$ and $w_b^\prime$ are indicated. Dark areas
    mark large spectral weight, the color scale is normalized in each
    plot separately.
  }
\end{figure}

Our results support the description of TiOCl as a layered two-dimensional
compound
with strong anisotropy, also on the level of
correlations. There is finite dispersion also perpendicular to the
chains, but a backfolding of the bands can only be seen along the
chains, where correlations are dominant.

Let us now comment on the relation of our results to 
ARPES data. Experiments show\cite{ho.si.05} 
that the dispersion in
TiOCl shows a strong asymmetric behavior along the
crystallographic $b$ direction, see also the right plot of \fig{comp_exp}. 
The binding energy of the lowest-lying band around $\mathbf
k=(0,0)$ is about $-1.5$\,eV, whereas around $\mathbf k=(0,\pi)$ it is
about $-2.0$\,eV. First attempts to describe the dispersions within
{\it ab initio} calculations were not successful. Standard LDA
calculations do not produce the backfolding of the bands induced by
short-range spin fluctuations, and spin-polarized LSDA+U calculations
cannot account for the asymmetry of the spectra.
Also the spectra of the
one-dimensional Hubbard model calculated within the dynamical
density-matrix renormalization group (DDMRG)
do not reproduce the experimental spectral weight
distribution.\cite{ho.si.05} However, as can be seen in \fig{comp_exp},
our new results show that the Hubbard model {\em can indeed} give a
good description of the asymmetry, since 
the inter-chain processes give rise to a spectral weight transfer
from the holon to the spinon band around the $\Gamma$ point, and
therefore make the band structure  more asymmetric. 

In order to compare our results more quantitatively, we extract the
following 
numbers related to the band widths of the spectra. The first one,
$w_b$, is the difference of the binding energies at $(0,\pi/2)$
and $(0,\pi)$ and is a measure for the overall band width. The second one,
$w_b^\prime$, is defined to be the difference in binding energies
between $(0,\pi/2)$ and $(0,0)$. The larger the difference between
these two quantities, the larger is the asymmetry of the
spectra. From 
experiments\cite{ho.si.07} we extract $w_b\approx 0.47$\,eV and
$w_b^\prime\approx 0.17$\,eV, and for the calculated spectra we get
$w_b\approx 0.50$\,eV and $w_b^\prime\approx 0.09$\,eV. Comparing
theory and experiment, we see that the overall band width $w_b$
is well reproduced by the calculation, but the asymmetry is even a bit
more pronounced in the theoretical spectra, resulting in a smaller
value of $w_b^\prime$. This result may be improved by
including more  longer-ranged hopping processes. Although they
decrease rapidly with distance, they can change the band widths within
a few percent. 

By inspecting \fig{comp_exp} it is obvious that the width of the
spectra is much larger in the ARPES data than in the calculated
spectra. The most important reason for that is that our
calculations are done at $T=0$, using exact diagonalization
techniques. Therefore there are no life-time effects due to finite
temperatures included in our calculations. Moreover, additional
coupling to lattice degrees of freedom could also lead to a smaller
life time, and hence broader excitations.

In summary, our results imply that it is important to include the
inter-chain couplings at least on a
single-particle non-interacting level into the effective model, in
order to improve the description of the experimental spectra, although
the strong correlations are constricted mainly to the chains.

With our work  we could fill the gap left by previous theoretical
studies regarding the momentum-resolved single-particle spectral function of TiOCl.
There are, though, still some open questions.
For example, we do see clear signatures of spin-charge separation in our
calculated spectra, which have not been found
experimentally. Also the so-called shadow band, dispersing at around
$-2$\,eV has not been seen in the ARPES spectra. 
A reason for this can be a very small relative
spectral weight of the high-energy band that
cannot be resolved in experiment. In
our calculation we also did not include the lattice degrees of
freedom, which are supposed to be very important in
 TiOCl\cite{pi.va.07}, driving the
spin-Peierls phase transition. These phonons can also lead to a
smearing of the peak structure of $A(\mathbf k,\omega)$.

Finally, we want to comment also on the differences between the
ARPES spectra of TiOCl and TiOBr. Experiments, supplemented with
band structure calculations, have shown\cite{le.ch.05,ho.si.07} that
in the latter compound 
the intra-chain couplings are  weaker and the inter-chain
couplings  stronger compared to TiOCl,\cite{le.ch.05}
e.\,g., $t_1$ decreases from $-0.21$\,eV to $-0.17$\,eV, whereas $t_3$
increases from $0.04$\,eV to $0.06$\,eV. By inspecting the
self-energies in a similar manner as we did in \fig{selfenergy},    
we found that also for these parameter values the
correlations are predominantly one-dimensional. There are only
changes in the overall bandwidths, but no qualitative changes. For
instance, the band width in $a$-direction is enhanced, but there are
no signatures of strong inter-chain correlations resulting in a
backfolding of the bands.

At this point we want to remark
that, in particular for TiOBr, one should be very careful with the
use of the spinon/holon terminology.
Although there are no qualitative changes due
to the enhanced couplings, they do change the band widths. Hence,
quantitative analysis have to be done including these inter-chain
couplings.

\section{Conclusions}\label{sec:discuss}

By combining {\it ab initio} calculations (LDA) and the variational cluster
approximation, we could study for the first time the
  momentum-resolved spectral function including a realistic band
  structure {\em and} strong-correlation effects.
In agreement with previous theoretical
studies and experimental results, our calculations showed an almost
complete polarization in the orbital sector, with 99\% of the
electrons occupying the Ti $d_{xy}$ orbital. 

Since the orbital degree of freedom is quenched, we could use an
effective single-band model for the investigation of the spectral
properties. The most striking result of our study is that the
inclusion of inter-chain hopping processes leads to a significant
spectral weight redistribution around the $\Gamma$ point from higher
to lower binding energies. This effect, which makes the spectrum more
asymmetric with respect to the points $(0,0)$ and $(0,\pi)$, cannot
be reproduced using only the hopping processes along the
chains. This result suggests that the frustrated inter-chain 
coupling\cite{zh.je.08}
is one of the main reasons for the strong asymmetry that has been
found in experimental ARPES measurements.  Moreover the calculated 
 spectral features may be extended  to a more general class of correlated
coupled-chains systems. 

An open question in TiOCl is still the role of phonons. Because of the vicinity
of the system to a spin-Peierls instability, the phonons are supposed
to be important in the system. Including these degrees of freedom,
although theoretically very demanding, 
could further improve the results 
with respect to lineshapes and lifetimes of the excitations.

\begin{acknowledgments}
This work has been supported by the Deutsche Forschungsgemeinschaft,
research unit 538, grant CL 124/6-1, the program SFB/TRR49, and the
Austrian Science Fund, grant J2760-N16. T.S.-D. thanks the
Max-Planck-Institue Stuttgart, Germany, through the partnergroup
program.
M.A. gratefully acknowledges useful discussions with E. Arrigoni,
M. Potthoff, and A. Georges. 
\end{acknowledgments}

\end{document}